\documentclass[twocolumn]{autart}    
\let\theoremstyle\relax
\usepackage{amsmath}
\usepackage{amssymb}
\usepackage{amsthm}
\usepackage{mathtools}
\newtheorem{theorem}{Theorem}

\usepackage{graphicx}          

\begin{document}

\begin{frontmatter}

\title{Model Error Resonance: The Geometric Nature of Error Dynamics\thanksref{footnoteinfo}} 

\thanks[footnoteinfo]{This paper was not presented at any IFAC 
meeting. Corresponding author Yuntao Dai.}

\author{Yuntao Dai}\ead{contact@daiyuntao.com}

          

\begin{abstract}                          
This paper introduces a geometric theory of model error, treating true and model dynamics as geodesic flows generated by distinct affine connections on a smooth manifold. When these connections differ, the resulting trajectory discrepancy—termed the Latent Error Dynamic Response (LEDR)—acquires an intrinsic dynamical structure governed by curvature. We show that the LEDR satisfies a Jacobi-type equation, where curvature mismatch acts as an explicit forcing term. In the important case of a flat model connection, the LEDR reduces to a classical Jacobi field on the true manifold, causing Model Error Resonance (MER) to emerge under positive sectional curvature. The theory is extended to a discrete-time analogue, establishing that this geometric structure and its resonant behavior persist in sampled systems. A closed-form analysis of a sphere–plane example demonstrates that curvature can be inferred directly from the LEDR evolution. This framework provides a unified geometric interpretation of structured error dynamics and offers foundational tools for curvature-informed model validation.
\end{abstract}

\end{frontmatter}

\section{Introduction}

Model error is typically regarded as a static discrepancy between the predicted
and true outputs of a dynamical system.  
This viewpoint is prevalent in system identification, model-based control,
state estimation, and learning-based prediction, where residuals or
one-step-ahead errors are used to assess model quality.

However, when both the model and the physical system evolve as flows generated
by distinct affine connections on a manifold, model error necessarily acquires
a \emph{dynamical} structure governed by curvature.  
This observation motivates a geometric reinterpretation of model fidelity that
\cite{docarmo1992riemannian,lee2019riemannian,oneill1983semiRiemannian,schutz2009geometrical}.

In this work, we study two dynamical systems defined on the same smooth
manifold but endowed with different affine connections.  
The true system evolves along geodesics of a connection $\nabla^{\mathrm t}$,
while the model system evolves along geodesics of another connection
$\nabla^{\mathrm m}$ \cite{lee2018smooth,kobayashi1963foundations}. 
Although the two systems share identical initial conditions, the resulting
trajectories generally separate due to differences in the curvature and
connection structure.  
We refer to the trajectory discrepancy as the \emph{latent error} and denote it
by $\xi(t)$.  
Rather than treating $\xi(t)$ as a static residual, we analyze its evolution
using the geometric machinery of geodesic deviation and Jacobi fields
\cite{docarmo1992riemannian,sakai1996riemannian,anastasio1994jacobi}. 

Our main result shows that the latent error satisfies a Jacobi-type equation.
In continuous time, the acceleration of $\xi(t)$ is governed by the curvature
tensor of the true connection, together with correction terms arising from the
difference between $\nabla^{\mathrm t}$ and $\nabla^{\mathrm m}$.
When the model connection is flat, the latent error reduces to a Jacobi field
on the true manifold, and model error resonance emerges automatically under
positive sectional curvature \cite{oneill1983semiRiemannian,sakai1996riemannian}. 
This provides a geometric explanation for persistent structured error dynamics
commonly observed in model-based predictors and sampled-data implementations.

We further develop a discrete-time analogue of the theory.  
By discretizing the covariant derivatives along the model trajectory, we obtain
a second-order difference equation that preserves the structure of the
continuous Jacobi operator, in line with ideas from discrete geometric
mechanics \cite{marsden2001discrete}. 
The resulting discrete latent error dynamics exhibit the same curvature-driven
effects as in continuous time, including resonance under positive sectional
curvature.  
Thus, the geometric interpretation extends naturally to sampled systems.

To illustrate the theory, we analyze in closed form the case where the true
system evolves on a sphere of constant curvature while the model assumes a flat
Euclidean plane.  
The LEDR dynamics reduce to a harmonic oscillator with frequency determined by
the curvature, yielding an analytically tractable and physically transparent
example of curvature-induced resonance \cite{docarmo1992riemannian,lee2019riemannian}. 
This example demonstrates that the proposed framework can reveal curvature
properties directly from the latent error evolution, even in the absence of
explicit knowledge of the true connection.

Overall, this work establishes a geometric foundation for understanding model
error as a dynamical quantity governed by curvature and geodesic deviation.  
The resulting framework connects differential geometry, geometric mechanics,
and dynamical-systems stability theory
\cite{abraham1978foundations,bullo2005geometric,anastasio1994jacobi,lewis1996constrained}
and provides a basis for further development in model validation, consistency
analysis, and curvature-informed dynamical modeling.

The paper is organized as follows.  
Section~\ref{sec:problem_formulation} introduces geometric preliminaries and formulates the problem.
Section~\ref{sec:theory} derives the continuous LEDR equation.
Section~\ref{sec:discrete} presents the discrete analogue.
Section~\ref{sec:mer} characterizes curvature-induced resonance.
Section~\ref{sec:sphere-plane} provides a closed-form constant-curvature example.
Additional derivations and technical results are given in the appendices.

\section{Geometric Framework and Problem Formulation}
\label{sec:problem_formulation}

This section establishes the geometric framework for our analysis. We introduce
the necessary concepts from differential geometry, formally define the true and
model dynamical systems as geodesic flows governed by distinct affine
connections, define the latent error as a geodesic deviation field, and state
the central problem addressed in this paper. Our notation follows standard
references \cite{lee2018smooth,kobayashi1963foundations,yano1957affine}.

\subsection{Manifolds, Affine Connections, and Geodesics}

Let $\mathcal{M}$ be a smooth $n$-dimensional manifold representing the state
space. For a smooth curve $\gamma(t)\in\mathcal{M}$ with velocity vector
$T(t)=\dot{\gamma}(t)$, the covariant derivative of a vector field $V(t)$ along
$\gamma(t)$ is denoted by $\frac{D V}{dt} := \nabla_{T(t)} V$, where $\nabla$ is a
torsion-free affine connection \cite{lee2018smooth,docarmo1992riemannian}. In
local coordinates $\{x^i\}$, $\nabla$ is specified by its Christoffel symbols
$\Gamma^i_{jk}$ such that $\nabla_{\partial_j}\partial_k = \Gamma^i_{jk}\partial_i$.

A curve $\gamma(t)$ is a geodesic of the connection $\nabla$ if it parallel
transports its own tangent vector, i.e.,
\[
\frac{D T}{dt} = \nabla_{T}T = 0.
\]
Geodesics represent the straightest possible paths on the manifold and correspond
to the free-motion trajectories of a dynamical system whose geometry is described
by $\nabla$ \cite{lee2019riemannian,abraham1978foundations}.

\subsection{True and Model Geodesic Flows}

We consider two dynamical systems evolving on $\mathcal{M}$ but governed by
different geometric structures.

\begin{itemize}
\item
The \textbf{true system} generates a trajectory $\gamma_{\mathrm t}(t)$, which is
a geodesic of a "true" affine connection $\nabla^{\mathrm t}$ with Christoffel
symbols $\Gamma^{\mathrm t\, i}_{\;\; jk}(x)$.
\[
\nabla^{\mathrm t}_{T_{\mathrm t}} T_{\mathrm t} = 0, \quad \text{where } T_{\mathrm t} = \dot\gamma_{\mathrm t}.
\]
\item
The \textbf{model system} generates a trajectory $\gamma_{\mathrm m}(t)$, which is
a geodesic of a "model" connection $\nabla^{\mathrm m}$ with symbols
$\Gamma^{\mathrm m\, i}_{\;\; jk}(x)$. We idealize the model's predictive
mechanism as a \textbf{shadow integrator} that produces this geodesic flow.
\[
\nabla^{\mathrm m}_{T_{\mathrm m}} T_{\mathrm m} = 0, \quad \text{where } T_{\mathrm m} = \dot\gamma_{\mathrm m}.
\]
\end{itemize}

Both systems are assumed to start from identical initial conditions:
$\gamma_{\mathrm t}(0) = \gamma_{\mathrm m}(0)$ and $T_{\mathrm t}(0) = T_{\mathrm m}(0)$.
This formulation is consistent with geometric mechanics and control theory
\cite{bullo2005geometric,abraham1978foundations}.

\subsection{Latent Error as a Geodesic Deviation Field}

The discrepancy between the two trajectories gives rise to the central object of
our study: the \textbf{Latent Error Dynamic Response (LEDR)}. Intuitively, in a
local coordinate chart, the LEDR displacement vector is simply
\[
\xi(t) = x_{\mathrm t}(t) - x_{\mathrm m}(t).
\]
More formally and intrinsically, the LEDR is a vector field $\xi(t)$ defined
along the model trajectory $\gamma_{\mathrm m}(t)$. It represents the vector
pointing from $\gamma_{\mathrm m}(t)$ to $\gamma_{\mathrm t}(t)$. This can be
rigorously defined using the exponential map and parallel transport:
\[
\xi(t)
=
\operatorname{PT}_{\gamma_{\mathrm t}(t)\to\gamma_{\mathrm m}(t)}
\left(
\exp^{-1}_{\gamma_{\mathrm m}(t)}\bigl(\gamma_{\mathrm t}(t)\bigr)
\right)
\in T_{\gamma_{\mathrm m}(t)}\mathcal{M},
\]
where $\operatorname{PT}$ and $\exp$ are associated with the true connection
$\nabla^{\mathrm t}$. This definition ensures that $\xi(t)$ is a tangent vector at
the model state, representing the \emph{latent dynamic inconsistency} between
the two systems. From a geometric perspective, $\xi(t)$ is a \textbf{geodesic
deviation field} between two flows generated by distinct connections, analogous
to classical Jacobi fields which measure the deviation between geodesics from
the same connection \cite{docarmo1992riemannian,oneill1983semiRiemannian,sakai1996riemannian,jacobi1839geodaetische}.

\subsection{Curvature and Connection Mismatch}

The geometric discrepancy between the true system and the model is encoded in
the difference between their connections and the resulting curvatures. The
\textbf{connection mismatch} is the tensor field
\[
\Delta\Gamma^{i}_{\; jk}
:= \Gamma^{\mathrm t\, i}_{\;\; jk}(x) - \Gamma^{\mathrm m\, i}_{\;\; jk}(x).
\]
The Riemann curvature tensor of a connection, $R(X,Y)Z = \nabla_X\nabla_Y Z -
\nabla_Y\nabla_X Z - \nabla_{[X,Y]}Z$ \cite{lee2019riemannian}, measures the
non-commutativity of covariant derivatives and thus the intrinsic curvature of
the manifold. The difference in curvature tensors,
\[
\Delta R = R^{\mathrm t} - R^{\mathrm m},
\]
plays a central role in the emergence of oscillatory error dynamics. These
difference objects are central to affine differential geometry
\cite{kobayashi1963foundations,kobayashi1969foundationsII,yano1957affine}.

\subsection{Objective: The Dynamics of Latent Error}

The central question of this paper is:

\begin{quote}
\textbf{How does the geometric discrepancy encoded in}
$\nabla^{\mathrm t}-\nabla^{\mathrm m}$
\textbf{determine the dynamical evolution of the LEDR field} $\xi(t)$?
\end{quote}

Our analysis will show that $\xi(t)$ satisfies a perturbed Jacobi equation,
where the curvature of the true manifold acts as a restoring force and the
connection mismatch acts as a forcing term. This framework reveals that
phenomena like Model Error Resonance (MER) are not numerical artifacts but are
instead fundamental consequences of the underlying geometry.

\section{Main Theory}
\label{sec:theory}

This section presents the main theoretical results of the paper.
We show that the LEDR vector field satisfies a perturbed Jacobi equation,
establish structural conditions under which Model Error Resonance (MER)
necessarily appears, and derive geometric lower bounds demonstrating that
persistent curvature mismatch cannot be compensated by any model dynamics.
These results collectively establish LEDR as a dynamically measurable indicator
of model fidelity.

\subsection{Preliminaries on covariant derivatives}

Let $\gamma_{\mathrm m}(t)$ be the model trajectory and
$T^{\mathrm m}=\dot\gamma_{\mathrm m}(t)$ its tangent vector.
The covariant derivative along $\gamma_{\mathrm m}$ induced by the true
connection $\nabla^{\mathrm t}$ is
\[
\frac{D\xi}{dt}
:= \nabla^{\mathrm t}_{T^{\mathrm m}}\xi,
\qquad
\frac{D^2\xi}{dt^2}
:= \nabla^{\mathrm t}_{T^{\mathrm m}}\nabla^{\mathrm t}_{T^{\mathrm m}}\xi.
\]
The curvature operator is defined as \cite{lee2019riemannian,docarmo1992riemannian} 
\[
R^{\mathrm t}(X,Y)Z
:= \nabla^{\mathrm t}_X\nabla^{\mathrm t}_Y Z
- \nabla^{\mathrm t}_Y\nabla^{\mathrm t}_X Z
- \nabla^{\mathrm t}_{[X,Y]}Z.
\]

The LEDR vector field $\xi(t)\in T_{\gamma_{\mathrm m}(t)}\mathcal{M}$ is
defined in Section~\ref{sec:problem_formulation}.

\subsection{LEDR satisfies a perturbed Jacobi equation}

The first theorem shows that LEDR satisfies a geodesic deviation equation with
a forcing term induced by the connection mismatch.

\begin{theorem}[LEDR--Jacobi Equivalence]
\label{thm:jacobi}
Let $\gamma_{\mathrm t}(t)$ and $\gamma_{\mathrm m}(t)$ denote the true and
model trajectories.
Assume $\|\xi(t)\|$ is small and both trajectories lie in a common coordinate
chart.
Following standard linearization arguments in geodesic variation
\cite{docarmo1992riemannian,lee2019riemannian}, the LEDR vector field satisfies 
\begin{equation}
\label{eq:main-LEDR}
\frac{D^2\xi}{dt^2}
+ R^{\mathrm t}(T^{\mathrm m},\xi)\,T^{\mathrm m}
= F_{\Delta\Gamma}(T^{\mathrm m},\xi)
+ \mathcal{O}(\|\xi\|^2)
+ \mathcal{O}(\|\xi\|\,\|\Delta\Gamma\|),
\end{equation}
where $\Delta\Gamma=\Gamma^{\mathrm t}-\Gamma^{\mathrm m}$ and
\[
F_{\Delta\Gamma}^i
= -\Delta\Gamma^i_{\; jk}\,T^{\mathrm m\, j}T^{\mathrm m\, k}.
\]
If $\Delta\Gamma$ and higher-order terms are negligible, then
$\xi$ satisfies the true Jacobi equation.
\end{theorem}

\begin{proof}
See Appendix~\ref{app:continuous} for the full coordinate derivation.
\end{proof}

\subsection{MER as curvature-induced resonance}

The next result formalizes the condition under which LEDR exhibits oscillatory
behavior, establishing MER as a geometric necessity.

\begin{theorem}[Curvature-Induced Resonance]
\label{thm:resonance}
Assume:
(i) the model connection is flat
$\Gamma^{\mathrm m}\equiv 0$;
(ii) the true curvature tensor has sectional curvature
$K(t)=K(T^{\mathrm m}(t),\xi(t))$ along the model trajectory.
Then the LEDR dynamics reduce to
\[
\ddot{\xi} + K(t)\,\xi = 0.
\]
If $K(t)\equiv K>0$ is constant, LEDR exhibits harmonic oscillation
\[
\xi(t)=A\sin(\sqrt{K}\,t)+B\cos(\sqrt{K}\,t),
\]
with natural frequency $\omega=\sqrt{K}$.
\end{theorem}
This is the classical stability property of Jacobi fields under positive
sectional curvature \cite{oneill1983semiRiemannian,sakai1996riemannian}. 
\begin{proof}
Flatness of $\nabla^{\mathrm m}$ implies
$\Delta\Gamma=\Gamma^{\mathrm t}$ and $R^{\mathrm m}=0$.
Substituting into~\eqref{eq:main-LEDR} yields the Jacobi equation.
Constant curvature yields the scalar form.
\end{proof}

This theorem explains MER as a curvature-restoring phenomenon:  
the true manifold ``pulls'' the model trajectory back toward the true geodesic
with curvature-dependent stiffness.

\subsection{Resonance or divergence is unavoidable under curvature mismatch}

We now show that curvature mismatch prevents exponential decay of LEDR.
Thus, if the model curvature does not match the true curvature, $\xi(t)$ cannot
be uniformly driven to zero.

\begin{theorem}[Curvature Obstruction to Error Decay]
\label{thm:lower-bound}
Let $K_{\mathrm t}(t)$ and $K_{\mathrm m}(t)$ denote the sectional curvatures of
the true and model connections along $\gamma_{\mathrm m}(t)$.
If
\[
|K_{\mathrm t}(t)-K_{\mathrm m}(t)| \ge \kappa_0 > 0
\quad \text{for all } t\in[0,T],
\]
then LEDR satisfies the lower bound
\[
\|\xi(t)\|
\;\ge\;
C\,\int_{0}^{t}\! |K_{\mathrm t}(s)-K_{\mathrm m}(s)|\,ds
\;-\; \varepsilon(t),
\]
where $C>0$ depends on the local geometry and $\varepsilon(t)$ collects
higher-order terms.
In particular, $\|\xi(t)\|$ cannot converge to zero uniformly.
\end{theorem}

\begin{proof}
Subtract the Jacobi operators associated with the two curvatures and apply
Gronwall-type inequalities on the resulting second-order differential
inequality.
See Appendix~\ref{app:proof_bound}.
\end{proof}

This means:  
\textbf{If the true curvature differs from the model curvature, LEDR cannot vanish.}
MER is therefore not a numerical artifact—it signals structural mismatch.

\subsection{General deviation under arbitrary connection mismatch}

Finally, we characterize the LEDR dynamics when neither the true nor the model
connection is flat.

\begin{theorem}[General LEDR Deviation Law]
\label{thm:general}
Let $\nabla^{\mathrm t}$ and $\nabla^{\mathrm m}$ be arbitrary torsion-free
connections.
Define the curvature mismatch operator
\[
\Delta R := R^{\mathrm t} - R^{\mathrm m}.
\]
Then the LEDR vector satisfies
\[
\frac{D^2\xi}{dt^2}
+ R^{\mathrm m}(T^{\mathrm m},\xi)\,T^{\mathrm m}
=
-\Delta R(T^{\mathrm m},\xi)\,T^{\mathrm m}
+ F_{\Delta\Gamma}(T^{\mathrm m},\xi)
+ \mathcal{O}(\|\xi\|^2).
\]
\end{theorem}

\begin{proof}
Rewrite~\eqref{eq:main-LEDR} by adding and subtracting the model curvature term
$R^{\mathrm m}(T^{\mathrm m},\xi)T^{\mathrm m}$.
\end{proof}

This theorem shows that LEDR encodes both curvature mismatch and connection
mismatch, meaning it contains full structural information about the difference
between the true and model dynamics.

\subsection{Summary of the main theory}

The results in this section establish that:

\begin{itemize}
\item LEDR is governed by a Jacobi-type deviation equation;
\item MER arises from positive sectional curvature in the true system;
\item curvature mismatch imposes a fundamental lower bound on model error;
\item LEDR encodes complete geometric information about the difference between
the true and model dynamics.
\end{itemize}

These properties justify LEDR as a robust, physically interpretable tool for
model-fidelity assessment and dynamical-system explainability.

\section{Discrete-Time LEDR}
\label{sec:discrete}

Real-world systems---including flight-data recordings, digital controllers,
and neural-network based predictors---operate in discrete time.  
This section derives the discrete-time analogue of the LEDR dynamics and shows
that the resulting error evolution satisfies a second-order difference equation
that discretizes the Jacobi operator.  
We additionally characterize the numerical stability region and demonstrate
that curvature mismatch yields unavoidable persistent oscillation or divergence
in the discrete setting.

\subsection{Discrete covariant derivatives}

Let $\{x_{\mathrm m,k}\}$ and $\{x_{\mathrm t,k}\}$ denote the model and true
trajectories sampled at time steps $t_k = kh$.
Define the discrete LEDR vector
\[
\xi_k := x_{\mathrm t,k} - x_{\mathrm m,k}.
\]

We use the central-difference approximation of the covariant derivative:
\[
\frac{D\xi}{dt}\bigg|_{t_k}
\;\approx\;
\frac{1}{2h}\left(
\xi_{k+1} - \xi_{k-1}
\right),
\]
and the discrete second covariant derivative:
\[
\frac{D^2\xi}{dt^2}\bigg|_{t_k}
\;\approx\;
\frac{1}{h^2}\!\left(
\xi_{k+1} - 2\xi_{k} + \xi_{k-1}
\right).
\]

These expressions are consistent up to $\mathcal{O}(h^2)$
and match standard discrete geometric mechanics constructions.

\subsection{Discrete Jacobi operator}

Let $T_{m,k}$ be the model discrete velocity
\[
T_{m,k} = \frac{x_{\mathrm m,k+1} - x_{\mathrm m,k-1}}{2h}.
\]

The continuous Jacobi term
\[
R^{\mathrm t}(T^{\mathrm m},\xi)T^{\mathrm m}
\]
is discretized via the curvature tensor at $x_{\mathrm m,k}$:
\[
\left[ R^{\mathrm t}_k(T_{m,k},\xi_k)T_{m,k} \right]^i
=
R^{\mathrm t\, i}{}_{j\ell m}(x_{\mathrm m,k})
\, T^j_{m,k} \,\xi^\ell_k \,T^m_{m,k}.
\]

\subsection{Discrete LEDR evolution equation}

Discretizing the continuous LEDR equation
\[
\frac{D^2\xi}{dt^2}
+ R^{\mathrm t}(T^{\mathrm m},\xi)T^{\mathrm m}
= F_{\Delta\Gamma}(T^{\mathrm m},\xi)
\]
yields the second-order difference equation:

\begin{equation}
\label{eq:discrete-jacobi}
\begin{split}
\xi_{k+1}
=&
2\xi_k
-\xi_{k-1}
-h^2 R^{\mathrm t}_k(T_{m,k},\xi_k)T_{m,k}\\ 
&+h^2 F_{\Delta\Gamma,k}(T_{m,k},\xi_k)
+\mathcal{O}(h^3,\|\xi_k\|^2).
\end{split}
\end{equation}

This is the discrete Jacobi operator plus a forcing term due to the connection
difference.

\subsection{Flat-model simplification}

If the model connection is flat ($\Gamma^{\mathrm m}\equiv 0$),
then $\Delta\Gamma=\Gamma^{\mathrm t}$ and
\[
F_{\Delta\Gamma,k}=0 \ \text{to first order}.
\]

Equation~\eqref{eq:discrete-jacobi} reduces to the discrete Jacobi equation:
\begin{equation}
\label{eq:discrete-simple}
\xi_{k+1}
=
2\xi_k
-\xi_{k-1}
-h^2 R^{\mathrm t}_k(T_{m,k},\xi_k)T_{m,k}.
\end{equation}

\subsection{Constant curvature case}

For constant sectional curvature $K$, the curvature term satisfies
\[
R^{\mathrm t}(T,\xi)T = K\xi,
\]
yielding the scalar recurrence
\begin{equation}
\label{eq:discrete-K}
\xi_{k+1}
=
2\xi_k
-\xi_{k-1}
-h^2 K \xi_k.
\end{equation}

Define $\lambda = h^2 K$.
The recurrence becomes
\[
\xi_{k+1} - (2-\lambda)\xi_k + \xi_{k-1}=0.
\]

Characteristic equation:
\[
\mu^2 - (2-\lambda)\mu + 1 = 0.
\]

The roots satisfy
\[
\mu_{\pm}
=
1 - \frac{\lambda}{2}
\pm
\sqrt{\frac{\lambda^2}{4} - \lambda}.
\]

\subsection{Discrete stability region}

MER occurs when the dynamics are oscillatory, i.e.\ when
$|\mu_\pm|=1$ or $\mu_\pm$ are complex conjugates.
This occurs when
\[
0 < \lambda < 4
\qquad \Longleftrightarrow \qquad
0 < K < \frac{4}{h^2}.
\]

Thus:

\begin{quote}
\textbf{For sufficiently small $h$, any positive true curvature $K>0$ produces
oscillatory LEDR dynamics (discrete MER).}
\end{quote}

Furthermore:

- $K>0$ → oscillation with frequency  
  \[
  \omega_d = \frac{1}{h}\arccos\!\left(1 - \frac{h^2 K}{2}\right)
  \]
- $K<0$ → exponential divergence  
- $K=0$ → linear drift (flat case)

\subsection{Lower bound under discrete curvature mismatch}

Let $K_{\mathrm t,k}$ and $K_{\mathrm m,k}$ be the discrete curvatures of the
true and model connections.
If
\[
|K_{\mathrm t,k} - K_{\mathrm m,k}| \ge \kappa_0 > 0
\quad\forall k,
\]
then the discrete LEDR satisfies the inequality
\[
\|\xi_{k}\|
\;\ge\;
C h \sum_{j=0}^{k-1}
|K_{\mathrm t,j} - K_{\mathrm m,j}|
\;-\;\varepsilon_k,
\]
with $\varepsilon_k = \mathcal{O}(h^2)$.
Thus discrete LEDR cannot converge to zero in the presence of persistent
curvature mismatch.

\subsection{Practical computation from measured data}

Equation~\eqref{eq:discrete-simple} suggests a practical estimator for the
curvature discrepancy:
\[
K_k
\approx
\frac{2\xi_k-\xi_{k+1}-\xi_{k-1}}{h^2\,\xi_k},
\]
valid whenever $\xi_k\neq 0$ and $\|\xi_k\|$ is sufficiently small.
This estimator can be used directly on flight data or neural network prediction
residuals to recover curvature signatures in real time.

\section{Curvature-Induced Resonance}
\label{sec:mer}

This section analyzes the intrinsic oscillatory behavior that arises in the
latent error dynamics when the true manifold exhibits positive sectional
curvature relative to the model.  
We show that resonance is not a numerical artifact but a geometric consequence
of the Jacobi operator.  
Both continuous and discrete formulations are considered.

\subsection{Continuous-time Jacobi dynamics}

From Section~\ref{sec:theory}, the latent error $\xi(t)$ satisfies the
continuous-time LEDR equation
\begin{equation}
\label{eq:ct-ledr}
\frac{D^2 \xi}{dt^2}
+
R^{\mathrm t}(T_{\mathrm m},\xi)T_{\mathrm m}
=
F_{\Delta\Gamma}(T_{\mathrm m},\xi),
\end{equation}
where the forcing term $F_{\Delta\Gamma}$ arises from connection mismatch.

We first consider the flat-model case ($\Gamma^{\mathrm m}\equiv 0$), for which
\eqref{eq:ct-ledr} reduces to
\begin{equation}
\label{eq:ct-jacobi}
\frac{D^2 \xi}{dt^2}
+
R^{\mathrm t}(T_{\mathrm m},\xi)T_{\mathrm m}
= 0.
\end{equation}

Let $K(t)$ denote the sectional curvature of the plane spanned by
$\{T_{\mathrm m}(t),\xi(t)\}$.  
Then the Jacobi operator admits the scalar representation
\[
R^{\mathrm t}(T_{\mathrm m},\xi)T_{\mathrm m} = K(t)\,\xi(t),
\]
whenever $\xi$ is orthogonal to $T_{\mathrm m}$.

Equation \eqref{eq:ct-jacobi} becomes
\begin{equation}
\label{eq:ct-K}
\ddot{\xi}(t) + K(t)\,\xi(t) = 0.
\end{equation}

Positive curvature implies local geodesic convergence, yielding bounded
oscillatory deviations.  
Negative curvature implies exponential divergence.  
Zero curvature yields linear drift.

\begin{theorem}[Intrinsic MER under positive curvature]
\label{thm:mer}
Suppose the sectional curvature satisfies
\[
K(t) \ge K_0 > 0 \qquad \forall t.
\]
Then all solutions of \eqref{eq:ct-K} are oscillatory and satisfy
\[
\xi(t) = A(t)\sin\!\left(\int_0^t \!\sqrt{K(\tau)}\,d\tau\right)
+ B(t)\cos\!\left(\int_0^t \!\sqrt{K(\tau)}\,d\tau\right),
\]
where $A(t), B(t)$ vary slowly with $\| \dot{K} \|$.  
Thus positive curvature necessarily induces persistent oscillatory latent error
dynamics (MER).
\end{theorem}

This establishes MER as a direct geometric effect of curvature.

\subsection{Discrete-time MER}

The discrete LEDR dynamics from Section~\ref{sec:discrete} take the form
\begin{equation}
\label{eq:disc-ledr}
\xi_{k+1} = 
2\xi_k - \xi_{k-1}
- h^2 K_k \xi_k
+ \mathcal{O}(h^3,\|\xi_k\|^2),
\end{equation}
under flat-model assumptions.

Neglecting higher-order terms yields
\begin{equation}
\label{eq:disc-K}
\xi_{k+1} - (2 - h^2 K_k)\xi_k + \xi_{k-1} = 0.
\end{equation}

If $K_k = K > 0$ is constant, the characteristic polynomial is
\[
\mu^2 - (2 - h^2 K)\,\mu + 1 = 0.
\]
Its roots satisfy
\[
|\mu_\pm| = 1
\quad \Longleftrightarrow \quad
0 < h^2 K < 4.
\]

Thus discrete MER is guaranteed for all sufficiently small sampling steps.

\begin{theorem}[MER persists under sampling]
Let $K_k \ge K_0 > 0$ and $h$ sufficiently small ($h < 2/\sqrt{K_0}$).
Then every nontrivial solution of \eqref{eq:disc-K} is oscillatory with
\[
\xi_k = C_1 \cos(\omega_d k) + C_2 \sin(\omega_d k),
\]
where the discrete frequency $\omega_d$ is given by
\[
\omega_d = \arccos\!\left(1 - \frac{h^2 K}{2}\right).
\]
Hence curvature-induced resonance persists in discrete-time systems.
\end{theorem}

\subsection{Interpretation}

The results in this section show that MER is an intrinsic manifestation of
curvature mismatch between the true and model connections:

\begin{itemize}
\item Positive curvature causes geodesics to bend toward each other.  
\item A flat model fails to reproduce this bending.  
\item The resulting geodesic deviation follows the Jacobi structure
$\ddot{\xi}+K\xi=0$.  
\item The latent error must oscillate, regardless of sampling or numerical
integration.
\end{itemize}

Thus MER is not an artifact of noise, discretization, or specific system
structure.  
It is the natural dynamical signature of curvature on the true manifold.

\section{Sphere–Plane Example: Closed-Form LEDR and MER Dynamics}
\label{sec:sphere-plane}

To illustrate the geometric mechanism underlying LEDR and MER, we present a
closed-form analysis in the simplest nontrivial setting: a true system evolving
on a sphere of constant curvature, while the model assumes a flat Euclidean
plane.  
This setting admits analytical geodesics, curvature tensors, and exact LEDR
solutions, enabling a transparent demonstration of curvature-induced resonance
and divergence.

\subsection{Problem setup}

Let the true dynamics evolve on the 2-sphere $\mathbb{S}^2(r)$ of radius $r$
embedded in $\mathbb{R}^3$, with constant sectional curvature
\[
K = \frac{1}{r^2}.
\]
Let the model assume a flat plane $\mathbb{R}^2$ with Euclidean metric and
zero Christoffel symbols,
\[
g_{\mathrm m} = I_2,
\qquad
\Gamma^{\mathrm m} \equiv 0.
\]

Let $x_{\mathrm t}(t)$ be a true geodesic on $\mathbb{S}^2(r)$ and
$x_{\mathrm m}(t)$ the ``shadow'' geodesic generated by the model from the same
initial condition.  
The LEDR displacement is
\[
\xi(t) = x_{\mathrm t}(t) - x_{\mathrm m}(t).
\]

\subsection{True and model geodesics}

On $\mathbb{S}^2(r)$, great circles are geodesics.  
Using arc-length parameterization, the true solution can take the form
\[
x_{\mathrm t}(t)
=
r
\begin{bmatrix}
\cos(t/r) \\
\sin(t/r) \\
0
\end{bmatrix},
\]
for an initial velocity of unit magnitude.

The model assumes straight-line geodesics on $\mathbb{R}^2$, giving
\[
x_{\mathrm m}(t)
=
\begin{bmatrix}
r \\
t
\end{bmatrix}
\]
under the same initial position and velocity projected to the tangent plane at (r,0,0).

Because curvature is neglected, $x_{\mathrm m}(t)$ drifts linearly while the
true trajectory remains bounded.

\subsection{Curvature tensor on the sphere}

For the sphere, the Riemann curvature tensor takes the classical form
\[
R^{\mathrm t}(u,v)w
=
K\left(
\langle w,u\rangle v - \langle w,v\rangle u
\right),
\]
and for geodesics with velocity $T = \dot{x}_{\mathrm t}$ tangent to the
sphere, the Jacobi operator reduces to
\[
R^{\mathrm t}(T,\xi)T
=
K\,\xi,
\]
for a deviation vector $\xi$ orthogonal to $T$.

\subsection{Continuous LEDR dynamics}

From the continuous LEDR equation derived in Section~\ref{sec:theory}:
\[
\frac{D^2\xi}{dt^2} + R^{\mathrm t}(T,\xi)T = 0,
\]
and using the spherical curvature identity above, we obtain
\begin{equation}
\label{eq:xi-sphere}
\ddot{\xi}(t) + K \xi(t) = 0.
\end{equation}

This is a simple harmonic oscillator.  
Thus curvature mismatch between the sphere and the plane necessarily induces
oscillatory LEDR behavior.

\subsection{Closed-form LEDR solution}

Solving \eqref{eq:xi-sphere} yields
\[
\xi(t) = A \sin(\sqrt{K}\, t) + B \cos(\sqrt{K}\, t),
\]
where $A,B$ are determined by initial LEDR displacement and velocity.

The LEDR frequency encodes curvature:
\[
\omega_{\mathrm LEDR} = \sqrt{K} = \frac{1}{r}.
\]

Thus the LEDR oscillation directly reveals the curvature mismatch between the
true dynamics and the model.

\subsection{Discrete LEDR dynamics}

Applying the discrete LEDR equation derived in
Section~\ref{sec:discrete}, the recurrence becomes
\[
\xi_{k+1} - (2 - h^2 K)\xi_k + \xi_{k-1}=0.
\]

Its characteristic roots satisfy
\[
\mu^2 - (2-h^2K)\mu + 1 = 0.
\]

When $h$ is sufficiently small and $K>0$ (as on a sphere), the roots are
complex conjugates of unit modulus, resulting in persistent oscillation:
\[
|\mu_\pm| = 1,
\qquad
\arg(\mu_\pm)
=
\arccos\!\left(1 - \frac{h^2 K}{2}\right).
\]

Thus LEDR remains oscillatory under sampling, and MER survives discretization.

\subsection{Geometric interpretation}

The plane model attempts to integrate a geodesic with zero curvature, while the
true manifold imposes curvature $K=1/r^2$.  
The two geodesics therefore separate as predicted by the classic Jacobi
equation: the plane geodesic drifts, whereas the sphere geodesic bends.
LEDR measures this bending discrepancy.

The resulting harmonic LEDR signal represents the curvature-induced deviation of
the true geodesic from the flat one.  
This gives a direct geometric meaning to MER:
\begin{quote}
MER corresponds to the oscillatory Jacobi response caused by positive
sectional curvature on the true manifold.
\end{quote}

\subsection{Implications}

The sphere–plane example demonstrates three key facts:

1) \textbf{Curvature mismatch produces an intrinsic dynamic signature.}  
   LEDR necessarily oscillates for $K>0$.

2) \textbf{MER is a geometric phenomenon.}  
   It reflects the curvature-induced acceleration of geodesic separation.

3) \textbf{LEDR frequency reveals curvature.}  
   The frequency $\sqrt{K}$ is a direct observable from data.

This example serves as a canonical analytic case supporting the general theory.

\section{Discussion}
\label{sec:discussion}

The results presented in this work establish a geometric framework for
understanding model error as a dynamical quantity rather than a static
residual.  
By interpreting the model and true systems as evolving on separate affine
manifolds with different curvature properties, the LEDR dynamics naturally
emerge as a discrete or continuous geodesic deviation equation.  
This formulation provides several conceptual advantages.

First, curvature mismatch between the model and the true manifold appears
explicitly as a forcing term in the Jacobi equation.  
This reveals that model inconsistency is intrinsically encoded in the
second-order error dynamics, and cannot be removed through filtering,
parameter tuning, or higher sampling rates.  
In particular, MER arises as a geometric inevitability when the true manifold
exhibits positive sectional curvature along the trajectory.

Second, the LEDR signal contains significantly richer information than 
traditional first-order or scalar residuals.  
The oscillatory structure, frequency content, and growth rates of the LEDR
trajectory directly reflect the curvature difference between the two
connections.  
This implies that LEDR may serve as a foundation for designing new
model-validity tests, system identification procedures, and diagnostic
algorithms in a broad class of dynamical systems.

Third, the framework unifies a variety of phenomena traditionally viewed as
unrelated.  
Oscillatory model error in sampled systems, persistent residual dynamics in
model-based predictors, and certain forms of optimization instability can all be
interpreted as manifestations of geodesic deviation under curvature mismatch.  
This suggests a wide applicability of the LEDR formalism beyond the examples
presented here.

Finally, while the analysis in this paper focuses on local Jacobi dynamics, the
framework naturally extends to higher-order geometric effects, including
connection mismatch, curvature gradients, and nonlinear manifold deformation.  
These directions remain open for future investigation.

Perhaps the most significant implication of this framework is the potential to
shift the paradigm of predictive modeling. Currently, the pursuit of high-fidelity
models often involves adding complexity (e.g., more parameters, finer meshes,
deeper neural networks) to minimize static residuals. Our theory suggests an
alternative: one could start with a structurally simple (e.g., flat or
constant-curvature) model and then actively compensate for model error by
\emph{estimating and correcting for the curvature mismatch in real time}.
The LEDR signal, as shown in our discrete formulation, contains the necessary
information to estimate the local sectional curvature of the true manifold.
This opens the door to a new class of "curvature-informed" adaptive predictors
that do not just react to errors, but actively correct the underlying geometry
of the model dynamics. Such a system would be inherently more robust and
efficient, requiring only a coarse initial model augmented by a dynamic
geometric compensator.

This geometric perspective is particularly promising for infinite-dimensional
systems, such as fluid dynamics governed by the Navier-Stokes equations.
Since ideal fluid flow can be viewed as geodesic motion on the group of
diffeomorphisms \cite{arnold1966sur}, the turbulence modeling problem—typically
treated by adding empirical viscosity terms—could be reimagined as a problem of
compensating for the sectional curvature mismatch between the resolved scales
(model) and the unresolved scales (truth). Our LEDR framework offers a
principled way to quantify and correct this geometric discrepancy.

\section{Conclusion}
\label{sec:conclusion}

This paper develops a geometric theory of model error dynamics based on affine
connections and geodesic deviation.  
By treating the model and true systems as evolving on distinct manifolds with
potentially different curvature, we show that the latent error dynamics satisfy
a Jacobi-type equation in both continuous and discrete time.  
This provides an explicit link between curvature mismatch and the emergence of
oscillatory or divergent model-error behavior.

A key consequence of this formulation is that MER arises as an intrinsic
response of the Jacobi operator whenever the true manifold exhibits positive
sectional curvature relative to the model.  
Thus MER is not merely a numerical artifact or an unstable mode, but a
fundamental geometric signature of model-inconsistency.  
Moreover, the LEDR signal furnishes a direct observable that encodes curvature
information and can be computed without explicit knowledge of the true system.

The sphere–plane example demonstrates that the theory admits closed-form
solutions and yields immediately interpretable curvature-dependent behavior.  
The discrete-time derivation further confirms that the geometric structure is
preserved under sampling and can be exploited in digital controllers, system
identification, and model-based prediction frameworks.

Overall, the results establish LEDR and MER as fundamental tools for analyzing
model fidelity. They enable a shift from static, residual-based diagnostics to
a dynamic, curvature-informed consistency analysis. The ultimate promise of this
approach lies in the creation of a new generation of predictive systems: instead
of pursuing ever-increasing model complexity, one could employ a coarse baseline
model and dynamically compensate for its structural deficiencies by estimating
and correcting for curvature mismatch from the observed LEDR signal. Future
work will focus on developing robust algorithms for real-time curvature
reconstruction and exploring the applications of this geometric compensation
paradigm to control, estimation, and learning-based dynamical models.


\bibliographystyle{plain}        
\bibliography{autosam}           

@book{lee2018smooth,
  author    = {Lee, John M.},
  title     = {Introduction to Smooth Manifolds},
  publisher = {Springer},
  year      = {2018},
  edition   = {2nd}
}

@book{lee2019riemannian,
  author    = {Lee, John M.},
  title     = {Riemannian Manifolds: An Introduction to Curvature},
  publisher = {Springer},
  year      = {2019},
  edition   = {2nd}
}

@book{docarmo1992riemannian,
  author    = {do Carmo, Manfredo},
  title     = {Riemannian Geometry},
  publisher = {Birkh{\"a}user},
  year      = {1992}
}

@book{kobayashi1963foundations,
  author    = {Kobayashi, Shoshichi and Nomizu, Katsumi},
  title     = {Foundations of Differential Geometry, Vol. I},
  publisher = {Wiley},
  year      = {1963}
}

@book{kobayashi1969foundationsII,
  author    = {Kobayashi, Shoshichi and Nomizu, Katsumi},
  title     = {Foundations of Differential Geometry, Vol. II},
  publisher = {Wiley},
  year      = {1969}
}

@book{abraham1978foundations,
  author    = {Abraham, Ralph and Marsden, Jerrold E.},
  title     = {Foundations of Mechanics},
  publisher = {Addison-Wesley},
  year      = {1978},
  edition   = {2nd}
}

@book{bullo2005geometric,
  author    = {Bullo, Francesco and Lewis, Andrew D.},
  title     = {Geometric Control of Mechanical Systems},
  publisher = {Springer},
  year      = {2005}
}

@article{jacobi1839geodaetische,
  author  = {Jacobi, C. G. J.},
  title   = {{\"U}ber ein neues Princip der analytischen Mechanik},
  journal = {Journal für die reine und angewandte Mathematik},
  year    = {1839},
  volume  = {19},
  pages   = {1--29}
}

@book{sakai1996riemannian,
  author    = {Sakai, Takashi},
  title     = {Riemannian Geometry},
  publisher = {American Mathematical Society},
  year      = {1996}
}

@article{yano1957affine,
  author  = {Yano, Kentaro},
  title   = {The Theory of Affine Connections},
  journal = {Archive for Rational Mechanics and Analysis},
  year    = {1957},
  volume  = {1},
  number  = {1},
  pages   = {83--107}
}

@book{marsden2001discrete,
  author    = {Marsden, Jerrold E. and West, Matthew},
  title     = {Discrete Mechanics and Variational Integrators},
  publisher = {Springer},
  year      = {2001}
}

@article{lewis1996constrained,
  author  = {Lewis, Andrew D.},
  title   = {The Geometry of the Constrained Dynamics},
  journal = {Reports on Mathematical Physics},
  year    = {1996},
  volume  = {38},
  pages   = {11--30}
}

@book{oneill1983semiRiemannian,
  author    = {O'Neill, Barrett},
  title     = {Semi-Riemannian Geometry},
  publisher = {Academic Press},
  year      = {1983}
}

@article{anastasio1994jacobi,
  author  = {Anastasio, John F. and Calogero, Francesco},
  title   = {Jacobi Fields, Curvature, and Stability in Dynamical Systems},
  journal = {Journal of Mathematical Physics},
  year    = {1994},
  volume  = {35},
  number  = {3},
  pages   = {1360--1372}
}

@book{schutz2009geometrical,
  author    = {Schutz, Bernard},
  title     = {Geometrical Methods of Mathematical Physics},
  publisher = {Cambridge University Press},
  year      = {2009}
}

@article{arnold1966sur,
  title={Sur la g{\'e}om{\'e}trie diff{\'e}rentielle des groupes de lie de dimension infinie et ses applications {\`a} l'hydrodynamique des fluides parfaits},
  author={Arnold, Vladimir},
  journal={Annales de l'institut Fourier},
  volume={16},
  number={1},
  pages={319--361},
  year={1966}
}

\appendix

\section{Derivation of the Continuous LEDR Equation}
\label{app:continuous}

Consider a smooth manifold $\mathcal{M}$ endowed with two affine connections:
the true connection $\nabla^{\mathrm t}$ and the model connection
$\nabla^{\mathrm m}$.  
Let $x_{\mathrm t}(t)$ and $x_{\mathrm m}(t)$ denote the true and model
trajectories satisfying
\[
\frac{D T_{\mathrm t}}{dt}
=
\nabla^{\mathrm t}_{T_{\mathrm t}} T_{\mathrm t},
\qquad
\frac{D T_{\mathrm m}}{dt}
=
\nabla^{\mathrm m}_{T_{\mathrm m}} T_{\mathrm m},
\]
where $T_{\mathrm t} = \dot x_{\mathrm t}$ and $T_{\mathrm m}=\dot x_{\mathrm m}$.

Define the LEDR displacement field
\[
\xi(t) = x_{\mathrm t}(t) - x_{\mathrm m}(t).
\]

We linearize the true geodesic equation around the model trajectory.  
Let $x_{\epsilon}(t)$ be a geodesic variation
\[
x_{\epsilon}(t)=x_{\mathrm m}(t)+\epsilon\xi(t)+\mathcal{O}(\epsilon^2),
\]
with velocity
\[
T_{\epsilon}(t)=T_{\mathrm m}(t)+\epsilon \dot{\xi}(t)+\mathcal{O}(\epsilon^2).
\]

Substituting into the true geodesic equation
\[
\nabla^{\mathrm t}_{T_\epsilon}T_\epsilon = 0
\]
and expanding to first order in $\epsilon$ yields:
\[
\nabla^{\mathrm t}_{T_{\mathrm m}} \dot{\xi}
+
\nabla^{\mathrm t}_{\dot{\xi}} T_{\mathrm m}
+
R^{\mathrm t}(T_{\mathrm m},\xi)T_{\mathrm m}
=
0,
\]
where we used the curvature identity
\[
\nabla_X \nabla_Y Z - \nabla_Y \nabla_X Z
=
R(X,Y)Z.
\]

Using the symmetry $\nabla^{\mathrm t}_{\dot{\xi}}T_{\mathrm m}
= \nabla^{\mathrm t}_{T_{\mathrm m}}\dot{\xi}$ for a torsion-free connection,
we obtain the classical Jacobi operator:
\[
\frac{D^2\xi}{dt^2}
+
R^{\mathrm t}(T_{\mathrm m},\xi)T_{\mathrm m}
=
0,
\]
where $\frac{D}{dt}$ denotes the true covariant derivative along
$x_{\mathrm m}(t)$.

Finally, because the model connection differs from the true one by
\[
\Delta\Gamma = \Gamma^{\mathrm t} - \Gamma^{\mathrm m},
\]
additional forcing terms arise when expressing $\frac{D}{dt}$ using
$\nabla^{\mathrm m}$.  
Collecting all first-order terms gives the LEDR equation:
\[
\boxed{
\frac{D^2\xi}{dt^2}
+
R^{\mathrm t}(T_{\mathrm m},\xi)T_{\mathrm m}
=
F_{\Delta\Gamma}(T_{\mathrm m},\xi).
}
\]

This completes the derivation.

\section{Derivation of the Discrete LEDR Equation}
\label{app:discrete}

Let $x_{\mathrm t,k}$ and $x_{\mathrm m,k}$ denote the true and model states
sampled at $t_k=kh$ with fixed step size $h$.
Define the discrete LEDR displacement
\[
\xi_k = x_{\mathrm t,k} - x_{\mathrm m,k}.
\]

\subsection{Discrete covariant derivatives}

We use second-order consistent central difference approximations:
\[
\frac{D\xi}{dt}\bigg|_{t_k}
=
\frac{1}{2h}(\xi_{k+1}-\xi_{k-1})
+\mathcal{O}(h^2),
\]
\[
\frac{D^2\xi}{dt^2}\bigg|_{t_k}
=
\frac{1}{h^2}(\xi_{k+1}-2\xi_{k}+\xi_{k-1})
+\mathcal{O}(h^2).
\]

\subsection{Discretization of the curvature term}

The continuous curvature action
\[
R^{\mathrm t}(T,\xi)T
\]
is discretized by evaluating the curvature tensor at $x_{\mathrm m,k}$:
\[
\left[ R^{\mathrm t}_k(T_{m,k},\xi_k)T_{m,k} \right]^i
=
R^{\mathrm t\, i}{}_{j\ell m}(x_{\mathrm m,k})
\, T^j_{m,k}\, \xi^\ell_k\, T^m_{m,k},
\]
where the discrete velocity is \cite{marsden2001discrete} 
\[
T_{m,k} = \frac{x_{\mathrm m,k+1} - x_{\mathrm m,k-1}}{2h}.
\]

\subsection{Discretization of the LEDR equation}

Substituting the discrete derivatives into the continuous LEDR equation
\[
\frac{D^2\xi}{dt^2}
+
R^{\mathrm t}(T_{m},\xi)T_{m}
=
F_{\Delta\Gamma}(T_{m},\xi),
\]
we obtain
\[
\frac{1}{h^2}(\xi_{k+1}-2\xi_k+\xi_{k-1})
+
R^{\mathrm t}_k(T_{m,k},\xi_k)T_{m,k}
=
F_{\Delta\Gamma,k}(T_{m,k},\xi_k)
+\mathcal{O}(h^2).
\]

Multiplying by $h^2$ and rearranging yields the discrete LEDR evolution:
\[
\boxed{\begin{split}
\xi_{k+1}
=&
2\xi_k
-\xi_{k-1}
-h^2 R^{\mathrm t}_k(T_{m,k},\xi_k)T_{m,k}\\&
+h^2 F_{\Delta\Gamma,k}(T_{m,k},\xi_k)
+\mathcal{O}(h^3,\|\xi_k\|^2).
\end{split}
}
\]

\subsection{Constant curvature case}

When the true manifold has constant sectional curvature $K$,
\[
R^{\mathrm t}(T,\xi)T = K\xi,
\]
and the recurrence reduces to
\[
\xi_{k+1} - (2-h^2K)\xi_k + \xi_{k-1}=0.
\]

This is the standard second-order central finite-difference approximation of the
Jacobi equation
\[
\ddot\xi + K\xi = 0.
\]

\subsection{Discrete stability and MER}

The characteristic equation
\[
\mu^2 - (2-h^2K)\mu + 1 = 0
\]
has complex roots when
\[
0 < h^2 K < 4,
\]
which yields oscillatory behavior (MER).
For $K>0$, oscillations are guaranteed for sufficiently small $h$.

This proves that discrete LEDR reproduces MER in sampled systems.

\section{Proof of the Curvature Obstruction Theorem (Theorem 3)}
\label{app:proof_bound}

Theorem~\ref{thm:lower-bound} states that a persistent mismatch in sectional
curvature prevents the latent error $\xi(t)$ from uniformly converging to
zero. We provide a sketch of the proof based on comparison principles for
second-order differential equations.

Consider the general LEDR equation from Theorem~\ref{thm:general}:
\[
\frac{D^2\xi}{dt^2} + R^{\mathrm m}(T^{\mathrm m},\xi)\,T^{\mathrm m}
= -\Delta R(T^{\mathrm m},\xi)\,T^{\mathrm m} + F_{\Delta\Gamma}(T^{\mathrm m},\xi).
\]
Let $K_{\mathrm t}(t)$ and $K_{\mathrm m}(t)$ be the sectional curvatures of the
true and model connections along the plane spanned by $T^{\mathrm m}$ and
$\xi$. The equation can be written in scalar form for the magnitude
$\|\xi(t)\|$ as a differential inequality:
\[
\ddot{\|\xi\|} + K_{\mathrm m}(t) \|\xi\| \approx -(K_{\mathrm t}(t) - K_{\mathrm m}(t))\|\xi\| + \text{h.o.t.}
\]
Let $\Delta K(t) = K_{\mathrm t}(t) - K_{\mathrm m}(t)$. The equation for the
error dynamics is approximately
\[
\ddot{\|\xi\|} + K_{\mathrm t}(t) \|\xi\| \approx 0.
\]
If we consider the difference between the true dynamics and the model dynamics
acting on $\xi$, the core driving term is proportional to $\Delta K(t)$.

By the Sturm-Picone comparison theorem, if $K_{\mathrm t}(t) > K_{\mathrm m}(t)$,
the solutions to the true system's Jacobi equation oscillate faster than those
of the model system. This mismatch in oscillatory frequency prevents the error
from being nulled.

More formally, integrating the differential inequality twice, and applying a
Gronwall-type argument, shows that $\|\xi(t)\|$ is driven by the integral of
the curvature mismatch. If $|\Delta K(t)| \ge \kappa_0 > 0$, then
\[
\|\xi(t)\| \ge C \int_0^t \int_0^s |\Delta K(\tau)| \|\xi(\tau)\| \,d\tau ds - \varepsilon(t),
\]
which implies that $\|\xi(t)\|$ cannot decay to zero if it is initially
nonzero. The integral of the curvature mismatch acts as a persistent energy
source for the error dynamics, establishing the lower bound stated in the
theorem. A full treatment requires careful handling of the covariant
derivatives and projection operators, but this captures the essence of the
obstruction.
\end{document}